\documentclass[printer]{aa}  

\usepackage{graphicx, xcolor}
\usepackage[varg]{txfonts}
\usepackage{upgreek}
\usepackage[normalem]{ulem}
\usepackage{comment}
\newcommand{\diff}{{\rm d}}

\DeclareRobustCommand{\uppartial}{\text{\rotatebox[origin=t]{20}{\scalebox{0.95}[1]{$\partial$}}}\hspace{-1pt}}

\newcommand{\ppardir}[2]{\ensuremath{\frac{\uppartial }{\uppartial #1} \left( #2\right)}}

\begin{document} 

\title{Mechanical model of a boundary layer for the parallel tracks of kilohertz quasi-periodic oscillations
in accreting neutron stars}

\titlerunning{QPOs from a neutron-star boundary layer}

\author{Pavel Abolmasov\inst{1,2}
\and
Juri Poutanen\inst{1,3,4}
}

\institute{
Department of Physics and Astronomy, FI-20014 University of Turku, Finland
\and
Sternberg Astronomical Institute, Moscow State University, Universitetsky pr. 13, 119234 Moscow, Russia
\and
Space Research Institute, Russian Academy of Sciences, Profsoyuznaya 84/32, 117997 Moscow, Russia
\and
Nordita, KTH Royal Institute of Technology and Stockholm University, Roslagstullsbacken 23, SE-10691 Stockholm, Sweden
}

\date{}

\abstract{
Kilohertz-scale quasi-periodic oscillations (kHz QPOs) are a distinct feature of the variability of neutron star low-mass X-ray binaries. 
Among all the variability modes, they are especially interesting as a probe for the innermost parts of the accretion flow, including the accretion boundary layer (BL) on the surface of the neutron star. 
All the existing models of kHz QPOs explain only part of their rich phenomenology. 
Here, we show that some of their properties may be explained by a very simple model of the BL that is spun up by accreting rapidly rotating matter from the disk and spun down by the interaction with the neutron star. 
In particular, if the characteristic time scales for the mass and the angular momentum transfer from the BL to the star are of the same order of magnitude, our model naturally reproduces the so-called parallel tracks effect, when the QPO frequency is correlated with luminosity at time scales of hours but becomes uncorrelated at time scales of days. 
 The closeness of the two time scales responsible for mass and angular momentum exchange between the BL and the star is an expected outcome of the radial structure of the BL.
}


 \maketitle

\section{Introduction}
\label{sec:introduction}
Stitching an accretion disk rotating at about Keplerian rate with the central object rotating much slower leads to the concept of accretion boundary layer (BL). 
The reason for talking about BL as an entity separate from the accretion disk is the inevitable breakdown of the basic assumptions of the standard disk theory in a very narrow region just above the surface of the accretor \citep{LP74, PS86}. 

In neutron star (NS) low-mass X-ray binaries (LMXBs), the BL is thought to be an important source of radiation, when the magnetic field of an accreting NS is too weak to support a magnetosphere. 
Shining at a luminosity comparable to that of the accretion disk \citep{LP74, SS00},  but being much more compact, BL is expected to have a harder spectrum and shorter variability time scales. 
Such a component has indeed been identified in LMXBs spectrally \citep{SP06, RSP13} as well as via its timing properties, in particular, as a source of kilohertz  quasi-periodic oscillations (kHz QPOs) \citep{GRM03}.
The position of the BL at the surface of the NS makes it a valuable probe for the fundamental properties of the star: its size, radius, and the physical conditions on its surface. 

The kHz QPOs have been observed in many NS LMXBs \citep{vdKlis_review}. 
Their frequencies span the range between about 200~Hz and the Keplerian frequency near the surface (about 1.3~kHz; see \citealt{mendez99,Belloni05}). 
Either one or two peaks, with the frequency difference of about 300~Hz \citep{mendez07}, are observed. 
In individual sources, kHz QPO frequencies may vary by a factor of 1.5--2. 
The frequencies are correlated with flux on time scales of hours  \citep{mendez99,mendez01}, while the correlation disappears on time scales of days. 
This phenomenon, known as QPO parallel tracks, was explained in a purely phenomenological way by \citet{vdKlis_parallel}.  
He suggested that the instantaneous X-ray luminosity of the source is a linear combination of the mass accretion rate $\dot{M}$ and its running average $\langle \dot{M}\rangle$, while the oscillation frequency is a function of $\dot{M}/\langle \dot{M}\rangle$ only. 
In this paper, we propose a mathematically similar but a more physically motivated solution to the parallel tracks problem. 

Here, we develop a simple mechanical model of the BL, which is treated as a thin massive belt supplied by the mass and angular momentum from the accretion disk and at the same time losing mass and angular momentum to the NS. 
We will consider the rotation frequency of the BL as the characteristic frequency responsible for kHz QPOs, though the real situation is probably much more complicated \citep[e.g.][]{ANP20}. 
In Sect.~\ref{sec:model}, we introduce the main equations based on conservation laws. 
In Sect.~\ref{sec:res}, we consider the properties of the model by solving the equations numerically. 
We discuss the results in Sect.~\ref{sec:disc}. 

\section{Model setup}\label{sec:model}

We consider the BL as an infinitely thin equatorial belt on the surface of a NS of radius $R$ and mass $M_{\rm NS}$ rotating at an angular frequency $\Omega_{\rm NS}$. 
Rotation of the layer is aligned both with the rotation of the star and the disk. 
Dynamics of the layer may be reduced to two equations, one for the mass and the other for the angular momentum conservation. 
The conservation law for the BL mass $M$ may be written as
\begin{equation}\label{E:dM}
\frac{\diff M}{\diff t} = \dot{M} - \frac{M}{t_{\rm depl}},
\end{equation}
where $\dot{M}$ is the mass supply rate from the disk. The second term describes mass precipitation from the BL onto the NS surface with the depletion time scale $t_{\rm depl}$ that exceeds the characteristic dynamical (Keplerian) time scale $t_{\rm dyn} = 1/\Omega_{\rm K} = \sqrt{R^3/GM_{\rm NS}}$, where $\Omega_{\rm K}$ is the Keplerian frequency.

Conservation of the angular momentum also involves sources and sinks related to the interaction with the surface of the star.
Hydrodynamic numerical simulations \citep[e.g.][]{belyaev13} suggest that the interaction between the BL and the surface of the star mediated by Reynolds stress is relatively weak. The relevant tangential stress  $W_{r\varphi} \sim 10^{-6}P$, where $P$ is the pressure at the bottom of the layer.
The impact of magnetic fields on the internal dynamics of the layer is probably important \citep{armitage02}, but it is unclear if they can provide an efficient angular momentum transfer between the BL and the star.
We will assume that the stress at the bottom of the BL is proportional to the pressure with a small proportionality coefficient $\alpha \ll 1$,
\begin{equation}\label{E:Rey}
W_{r\varphi} = \alpha P = \alpha g_{\rm eff} \Sigma,
\end{equation}
where $\Sigma$ is the BL surface density and 
\begin{equation}\label{E:geff}
g_{\rm eff} = \frac{GM_{\rm NS}}{R^2} - \Omega^2 R
\end{equation}
is the effective surface gravity, where $\Omega$ is the rotation frequency of the layer. This allows to express the braking torque acting on the layer as
\begin{equation}\label{E:Tminus}
T^- = A R W_{r\varphi} =  \alpha g_{\rm eff} M R,
\end{equation}
where $A$ is the surface area of the BL (projected onto the surface of the star) and the BL mass is $M=A\Sigma$. 

The angular momentum conservation law including mass depletion and friction takes the form
\begin{equation}\label{E:dJ}
  \frac{\diff J}{\diff t} = \dot{M}j_{\rm d} - \frac{J}{t_{\rm depl}} - \alpha g_{\rm eff} M R,
\end{equation}
where $J = \Omega M R^2$ is the total angular momentum of the layer, $j_{\rm d} = \sqrt{GM_{\rm NS}R}$ is the specific angular momentum of the matter entering from the disk.
We ignore viscous interaction between the disk and the BL. This corresponds to the ``accretion gap'' scenario \citep{KW85} when the last stable orbit is located above the surface of the NS, and thus the disk is causally disconnected from the BL. 
Recent constraints for the NS radius \citep{nattila17,MLD_nicer19, RWB_nicer19,capano20} suggest that this should be the case, at least below the Eddington limit.

Two equations (\ref{E:dM}) and (\ref{E:dJ}) are sufficient to describe the evolution of the physical parameters of the BL with time, given $\dot{M}(t)$ and initial conditions.
In our framework, the energy released during accretion and dissipation does not affect the dynamics of the layer. 
However, luminosity is an important observable. 
Some of the kinetic energy of the flow contributes to the spin-up of the star and the rest is converted to heat and contributes to the luminosity. 
The dissipated luminosity may be found as the change in the kinetic energy (see e.g. Appendix B of \citealt{PN95}). 
Our model splits this spin-down of the gas being accreted into two episodes: some dissipation occurs when the matter from the disk enters the BL at the rate $\dot{M}$, and some during the matter depletion from the BL (at the rate of $M/t_{\rm depl}$). 
In addition to these two components, there is viscous dissipation unrelated to mass exchange, equal to one half of the stress $W_{r\varphi}$ times the strain $R\diff\Omega/\diff R$ (see \citealt[section 16]{LL_hydro}). 
Together, the luminosity associated with the BL may be written as the sum of three terms
\begin{eqnarray}\label{E:Ltot}
 \displaystyle L & = & 
\displaystyle  \frac{1}{2}\dot{M} R^2 \left( \Omega_{\rm d}^2 - \Omega^2\right) + \frac{1}{2} \alpha g_{\rm eff} MR \left( \Omega - \Omega_{\rm NS}\right) \nonumber \\
& + & \frac{1}{2} \frac{M}{t_{\rm depl}} R^2 \left( \Omega^2 -\Omega_{\rm NS}^2\right). 
\end{eqnarray}
The first term on the right-hand side is the kinetic energy lost by the matter that enters the BL from the disk with the angular frequency $\Omega_{\rm d} = j_{\rm d}/R^2$. 
The second term is the viscous dissipation associated with the Reynolds stress~(\ref{E:Rey}). 
The last term corresponds to the kinetic energy of the BL material that precipitates onto the NS and acquires its rotation velocity. 

Below, we will assume that the BL is fed by a variable source of mass. We will assume stochastic variability of the mass accretion rate, modeled as a white noise source convolved with a kernel corresponding to a power-law power-density spectrum (PDS) with a random Fourier image phase (that corresponds to a random moment in time and unsynchronized variability at different frequencies). 
Integrating white noise leads (as it involves summation of a large number of independent random numbers) to a normally distributed quantity. 
To reproduce the log-normal flux distribution reported in many observational works \citep{uttley05}, we then exponentiate the result of the convolution and re-normalize it to match the mean value of $\dot{M}$. 

\section{Results}\label{sec:res}

\subsection{Approach to the equilibrium solution}

For a fixed BL mass and mass accretion rate, rotation of the BL may be described in terms of approach to a single equilibrium state. 
Using Eqs.~(\ref{E:dM}) and (\ref{E:dJ}), we can derive an evolutionary equation for $\Omega$: 
\begin{eqnarray}
  \displaystyle   \frac{\diff \Omega}{\diff t} &=& \frac{\diff}{\diff t}\left( \frac{J}{M R^2}\right) = \frac{J}{MR^2} \left( \frac{\dot{M}j_{\rm d}}{J} - \frac{\dot{M}}{M} - \frac{\alpha g_{\rm eff}MR}{J}\right) \nonumber \\
  \displaystyle &=& \frac{\dot{M}}{M}\left( \Omega_{\rm d} - \Omega\right) - \alpha \left( \Omega_{\rm K}^2 - \Omega^2\right).
\end{eqnarray}
The right-hand side of this equation is quadratic in $\Omega$, that allows to re-write it in the form
\begin{equation}\label{E:dO}
    \frac{\diff \Omega}{\diff t} = \alpha \left( \Omega_- - \Omega \right) \left( \Omega_+ - \Omega \right),
\end{equation}
where 
\begin{equation}\label{E:Opm}
    \Omega_{\pm} = \frac{\dot{M}}{2\alpha M} \pm  \sqrt{\left(\frac{\dot{M}}{2\alpha M}-\Omega_{\rm K}\right)^2 + \frac{\dot{M}}{\alpha M} \left(\Omega_{\rm K}-\Omega_{\rm d}\right)}.
\end{equation}
For $\Omega_{\rm d} = \Omega_{\rm K}$, one of the frequencies becomes $\Omega_+ = \Omega_{\rm K}$, and the other $\Omega_- = \dot{M}/(\alpha M) - \Omega_{\rm K}$.
The lower of the two roots, that is always $\Omega_-$ for the parameter values we consider (see Sect.~\ref{sec:omega} for more details), is stable. 

Our approximation is valid only if $\Omega < \Omega_{\rm K}$, otherwise effective gravity becomes negative and the flow is unbound. 
Unless $\Omega_-$ becomes smaller than $\Omega_{\rm NS}$, BL will evolve towards this equilibrium state. 
Otherwise, the layer stalls at $\Omega = \Omega_{\rm NS}$, and $W_{r\varphi}$ works as static friction. 

Mass equilibrium is reached when 
\begin{equation}\label{E:masseq}
M = M_{\rm eq} = \dot{M}t_{\rm depl}.
\end{equation}
When, at a fixed mass accretion rate, the system reaches both equilibrium mass and rotation frequency, the position of the stable stationary point depends, apart from $\Omega_{\rm d}/ \Omega_{\rm K}$ that we fix to 1, on a single parameter $\alpha M_{\rm eq} / \dot{M}$. 
It is easy to check that this quantity, multiplied by Keplerian frequency, is equal to the ratio of the characteristic depletion and friction time scales, 
\begin{equation}\label{E:q}
    q = \frac{t_{\rm depl}}{t_{\rm fric}} = \alpha \Omega_{\rm K} t_{\rm depl}.
\end{equation}
For $\Omega_{\rm d} = \Omega_{\rm K}$, the equilibrium rotation frequency is
\begin{equation}\label{E:Oeq}
    \Omega_{\rm eq} =  \left( \frac{1}{q}-1\right) \Omega_{\rm K}.
\end{equation}
When the friction becomes more efficient than depletion, the layer brakes down to $\Omega = \Omega_{\rm NS}$, that leads to trivial rotational evolution.
Hence, in the simulations with variable mass accretion rate, we will keep $\alpha \lesssim 1 /(\Omega_{\rm K} t_{\rm depl})$.

\subsection{Variable mass accretion rate}

If the mass accretion inflow to the layer is variable, the BL works as a filter for the variability of $\dot{M}$. 
The system of equations we consider is practically linear, though there is non-linearity introduced by $g_{\rm eff}$ in the friction term in Eq.~(\ref{E:dJ}). 
The characteristic depletion and friction time scales are presumably much longer than the dynamical time, and probably also exceed the viscous time scales in the inner disk. 
The outer disk, however, evolves even slower. 
In the relevant frequency range, the shapes of the PDSs of LMXBs are generally close to a power law ${\rm PDS} \propto f^{-p}$ with the slope of $p \simeq 1.3$ \citep{GA05}. 
We use this spectral slope in our simulations as representative of the variability of the disk. 

The mean mass accretion rate was set to Eddington $\dot{M} = L_{\rm Edd}/c^2$. 
The exact value does not affect the qualitative picture of accretion but sets the accretion time scale and equilibrium mass of the layer. 
As it was mentioned in Sect.~\ref{sec:model}, the variations of the mass accretion rate logarithm were considered as an integral of a white noise process. This allows to introduce one extra parameter, the dispersion of $\ln \dot{M}$. 
 In our simulations, we set the root-mean-square deviation of mass accretion rate logarithm $D = \sqrt{\left\langle\left(\Delta \ln\dot{M}\right)^2\right\rangle}$ to $0.5$.
This value allows to reproduce the relative variations of the characteristic frequencies without strong inconsistency with flux variation amplitudes in LMXBs \citep{HK89, mendez99}.

\begin{figure}
\centering
\includegraphics[width=0.85\columnwidth]{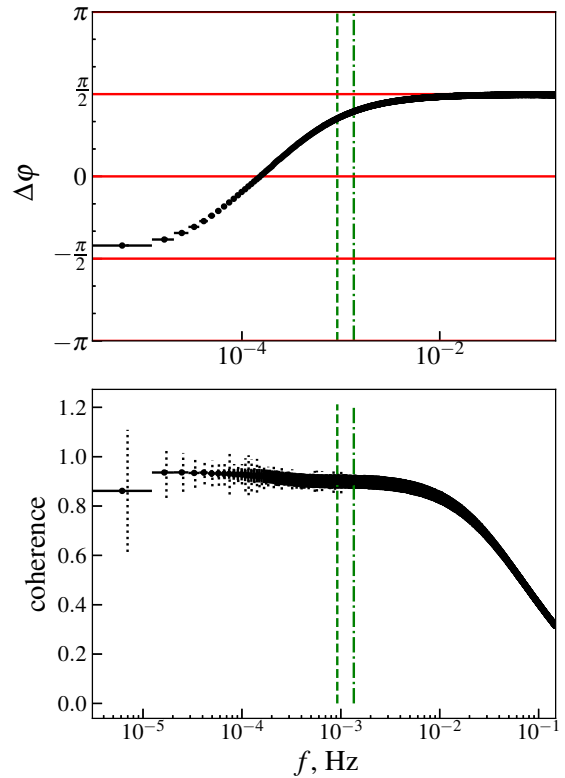}
 \caption{Phase lags (upper panel) and coherence (lower panel) between instantaneous luminosity of the BL $L$ and its rotation frequency $\Omega$. 
 Positive phase lag means that $\Omega$ lags $L$. 
 The green vertical lines show the frequencies corresponding to the depletion $t_{\rm depl}$ (dot-dashed) and the friction $1/\alpha\Omega_{\rm K}$ (dashed) time scales. Additional error bars (vertical dotted) show variability of the quantities within the bin. 
 The parameters are $\alpha = 10^{-7}$, $t_{\rm depl} \simeq 740$~s (corresponding to $q \simeq 0.68$), NS spin period of 3~ms. 
 }\label{fig:crossp}       
\end{figure}

In our model, the BL does not have any variability of its own, hence the variations of its luminosity are essentially smaller than that of the mass accretion rate, especially at high frequencies. 
In reality, of course, there is an additional variability component originating in the layer. 
The BL light curve is smoother and lags the mass accretion rate, as one would expect from the properties of the model where the BL emission depends on the history of mass accretion rate.

\begin{figure*}
\centering 
\includegraphics[width=1.\textwidth]{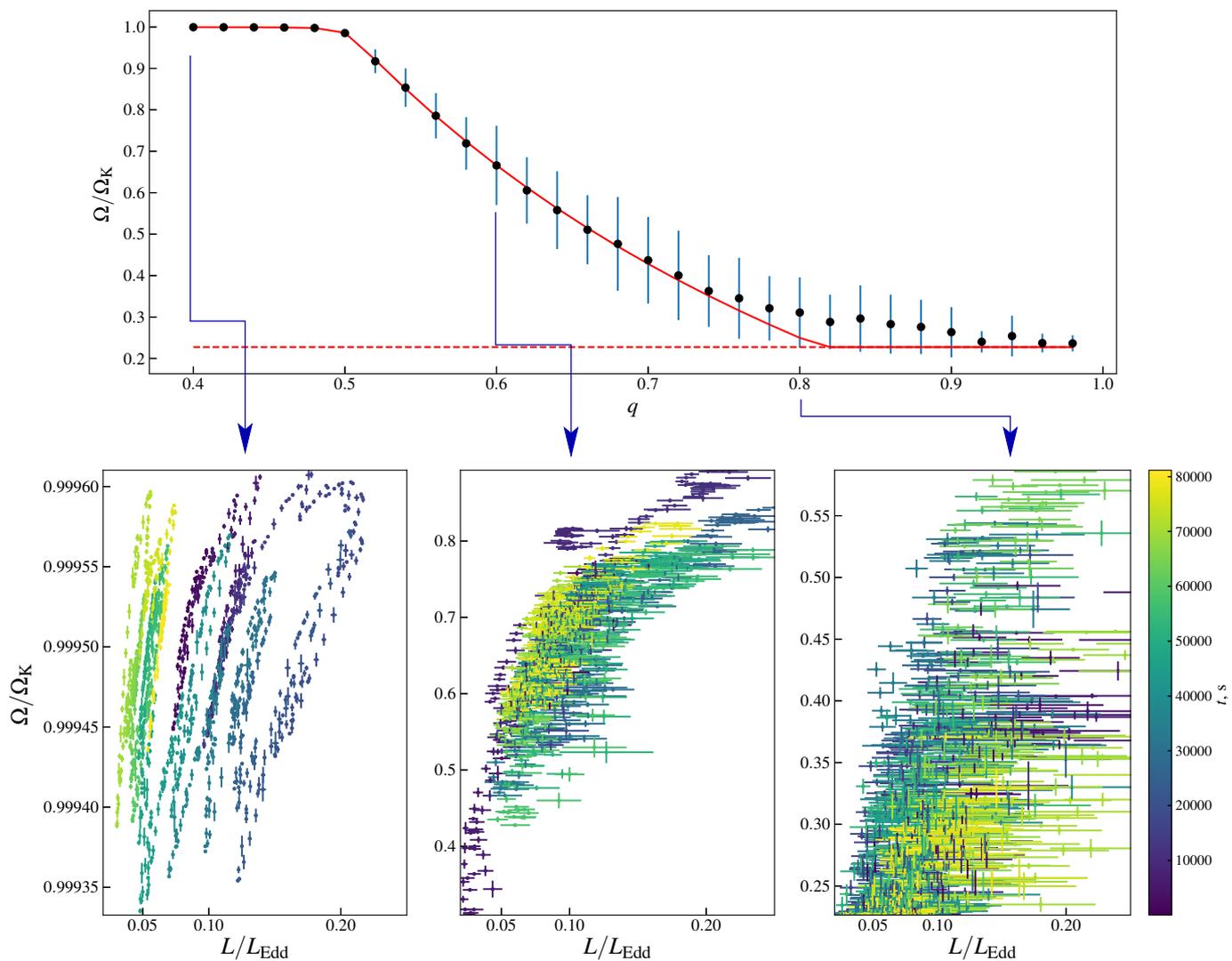} 
 \caption{
\textit{Upper panel}: mean BL rotation frequency (black dots; error bars show the root mean square variations of $\Omega$) as a function of $q$ compared to the equilibrium value $\Omega_{-}$ (red solid line). Dashed red line shows the rotation frequency of the NS (3ms).
\textit{Lower panels}: BL rotation frequency dependence on instantaneous luminosity for sample light curves with three different values of depletion time corresponding to $q= 0.4,$ $0.6$, and $0.8$. 
In all the simulations, $\alpha = 10^{-7}$. 
Time in the lower panels is color-coded, see the colorbar on the right.
The crosses in the lower panels are the average values calculated for 64~s-long time bins, the error bars show standard deviation. 
  }\label{fig:ofreq}       
\end{figure*}

We computed the cross-spectra of BL luminosity $L$ (see Eq.~\ref{E:Ltot}) and its rotation frequency $\Omega$ which are the proxies for the flux and QPO frequency, respectively. 
The argument of the cross-spectrum gives the phase lags, which we show in Fig.~\ref{fig:crossp} as a function of Fourier frequency. 
We also computed the coherence \citep{VN97, nowak99} shown in the lower panel of the figure. 
Both are averaged over a series of $10^4$ light curves. 

Quite expectedly, the quantities are correlated at lower frequencies but uncorrelated at $f\gg 1/t_{\rm depl, \, fric}$. 
Maximal coherence, however, occurs at intermediate frequencies $f\sim (0.1-1)/t_{\rm depl, \, fric}$. 
At higher frequencies, luminosity becomes sensitive to rapid variations in $\dot{M}$, uncorrelated with $\Omega$. 
Phase lags at low frequencies are negative, as the variations of $L$ lag the variations of $\dot{M}$, while $\Omega$ follows the variations of $\Omega^-(\dot{M}, M)$ (see Sect.~\ref{sec:omega}).
The phase lags increase with frequency and become positive at the time scales somewhat longer than the time scales of the BL ($\sim t_{\rm depl}$ and $t_{\rm fric}$). At high frequencies, they approach $\Delta \varphi = \uppi/2$.
Such a flat phase lag spectrum is a natural outcome of the mathematical properties of the initial system of equations.
The luminosity given by Eq.~(\ref{E:Ltot}) contains one term proportional to $\dot{M}$ (first term, related to the variable mass inflow to the BL). 
The other two terms depend only on $M = \int \dot{M} dt + \mbox{const}$ and on $\Omega$.
The spectral slope of $\dot{M}$ is always shallower than that of $M$. 
At a given frequency $f$, the friction and depletion terms have contributions $\sim 1/(ft_{\rm fric, \, depl})$ with respect to the first term.
Thus at high frequencies, variability of $L$ is dominated by  variations of the mass accretion rate.
Rotation frequency at high $f$ (when $M\simeq \mbox{const}$) is a result of integration of $\Omega_-$ (see Eq.~\ref{E:Opm}) that is a function of $\dot{M}$ and $M$. Taking Fourier transform of Eq.~(\ref{E:dO}) in the high-frequency limit yields
\begin{equation}
    2\uppi {\rm i} f \tilde{\Omega} \simeq \left(\Omega_{\rm K} - \Omega\right) \frac{\tilde{\dot{M}}}{M},
\end{equation}
where all the higher-order terms in $f$ are neglected, $\Omega_+$ replaced with $\Omega_{\rm K}$, and the Fourier transform of $\Omega_-$ replaced by $\tilde{\dot{M}}/\alpha M$.
Hence, in this limit, the Fourier image of rotation frequency is
\begin{equation}
    \tilde{\Omega} \simeq \frac{1}{2\uppi {\rm i} f} \frac{\tilde{\dot{M}}}{M} \left( \Omega_{\rm K}-\Omega\right).
\end{equation}
As $L$ is mainly affected by the first term, the cross-spectrum becomes
\begin{eqnarray}\label{E:CLO}
\displaystyle    C(L, \Omega) & \simeq & \frac{1}{2} R^2 \left( \Omega_{\rm K}^2-\Omega^2\right) \tilde{\dot{M}} \tilde{\Omega}^* \nonumber \\
\displaystyle &\simeq& \frac{\rm i}{4\uppi f} R^2 \left( \Omega_{\rm K}^2-\Omega^2\right)\left( \Omega_{\rm K}-\Omega\right) \left|\tilde{\dot{M}}^*\right|^2.
\end{eqnarray}
The argument of this expression is $\uppi/2$.

\subsection{Rotation frequency variations}\label{sec:omega}

Behavior of the BL, including its rotation frequency, depends strongly on the balance between mass and angular momentum loss, that may be described by the dimensionless quantity $q$ (see Eq.~\ref{E:q}). 
In Fig.~\ref{fig:ofreq} we show the mean rotation frequency and its variations for different values of $q$. 
Apparently, the mean value is well predicted by $\Omega_{-}$ given by Eq.~(\ref{E:Opm}).
When the depletion time scale is much shorter ($q\lesssim 0.5$), the BL co-rotates with the disk. 
In the opposite limit, friction spins the BL down to $\Omega_{\rm NS}$.
Strong variations in $\Omega$ are present only when the two time scales (friction and depletion) are comparable ($q \simeq 0.5-0.9$). 

\begin{figure}
\centering 
\includegraphics[width=0.9\columnwidth]{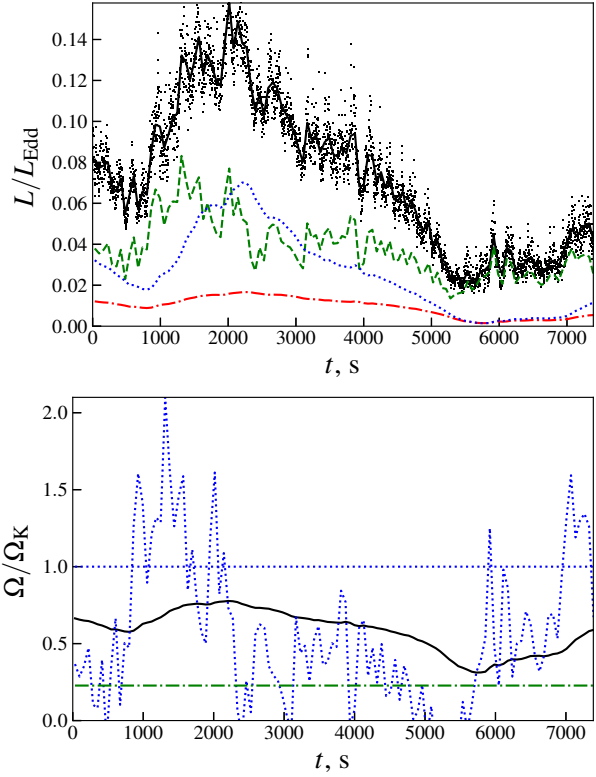}
 \caption{
\emph{Upper panel}: portion of the light curve of the simulation with  $\alpha = 10^{-7}$ and $q=0.6$. Solid black curve shows the total 64\,s-averaged luminosity (Eq.~\ref{E:Ltot}). We also show three contributions to the luminosity separately: first, second, and third terms from Eq.~(\ref{E:Ltot}) are plotted with green dashed, blue dotted, and red dot-dashed lines. Black dots are instantaneous luminosity values (every 2\,s). \emph{Lower panel}: rotation frequency $\Omega$ (black solid) and $\Omega_{\pm}$ (blue dotted) for the same model. 
Green dashed horizontal line corresponds to the spin of the NS. 
}\label{fig:olcurves}       
\end{figure}

\begin{figure*}
\centering 
\includegraphics[width=0.78\textwidth]{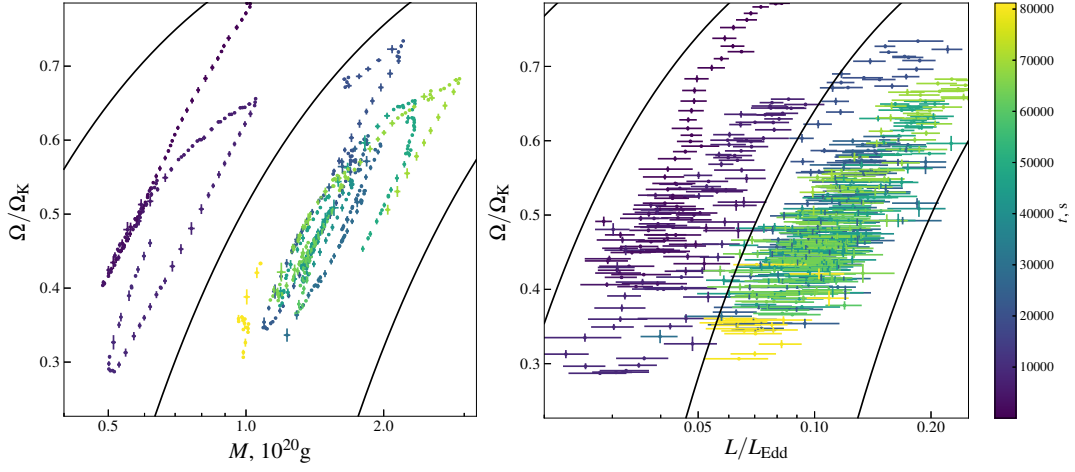}
 \caption{
Parallel tracks on the $M-\Omega$ and $L-\Omega$ planes for a simulation with $\alpha = 10^{-7}$, $q=0.56$, $D=0.5$, and $p=1.3$. Solid black lines are the lines of $\left(\Omega_{\rm K}-\Omega\right)M = $const and $\left(\Omega_{\rm K}-\Omega\right)L = $const. Time is color-coded (see the color bar on the right). To demonstrate the parallel tracks effect during multiple observation runs, we show only the data points in $10^4$\,s intervals separated by $10^4$\,s gaps.
}\label{fig:OM}       
\end{figure*}

In general, relation between the observed luminosity and rotation frequency of the layer is non-unique, and we qualitatively reproduce a parallel tracks picture (see the lower panels of Fig.~\ref{fig:ofreq}). 
On the shortest time scales much smaller than $t_{\rm depl, \, fric}$, variability of the luminosity is dominated by the first term in Eq.~(\ref{E:Ltot}), uncorrelated with $\Omega$. 
However, if the luminosity is averaged in time bins several times smaller than the time scales of the BL, it becomes correlated with $\Omega$.  
On these time scales, variations of $\Omega_{-}$ in Eq.~(\ref{E:dO}) dominate over variations of $\Omega$ (see Fig.~\ref{fig:olcurves}), hence rotation frequency derivative
\begin{equation}
    \frac{\diff \Omega}{\diff t} \simeq \left( \Omega_{\rm K} - \Omega \right) \frac{\dot M}{M}.
\end{equation}
Neglecting mass depletion, this yields
\begin{equation}
    M \propto \frac{1}{\Omega_{\rm K}-\Omega},
\end{equation}
where the proportionality coefficient is a slowly variable function of time. 
This is a scaling well reproduced in the evolution of the BL on the time scales several times smaller than friction and depletion scales (Fig.~\ref{fig:OM}). 
Luminosity variations also follow a similar trend $L \propto \left(\Omega_{\rm K}-\Omega\right)^{-1}$. 

\subsection{The influence of the other model parameters}

In spite of its simplicity, the model has several parameters, the values of which are not derived from the basic principles. 
The influence of the rotation frequency of the star $\Omega_{\rm NS}$ does not change the overall behavior. 
For the solutions with $q\lesssim 1$, it only limits the possible values of $\Omega$ and slightly modulates the spin-down term. 
The mean mass accretion rate in the framework of our model also plays a secondary role, affecting only the luminosity of the BL. 

\begin{figure*}
\centering 
\includegraphics[width=0.83\textwidth]{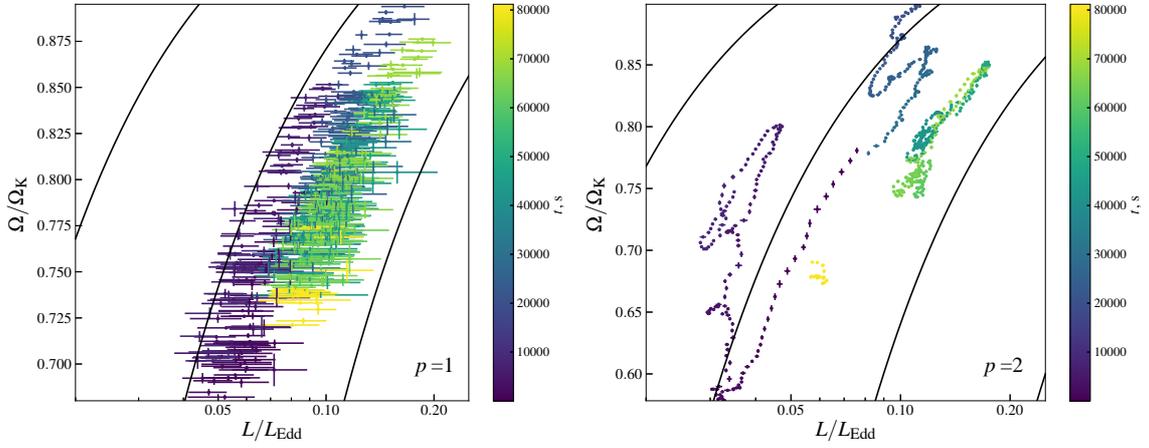}
 \caption{
Same as right panel of Fig.\,\ref{fig:OM} but for 
$p=1$ ({\it left panel}) and $p=2$ ({\it right panel}). 
}\label{fig:varp}       
\end{figure*}

\begin{figure*}
\centering 
\includegraphics[width=0.83\textwidth]{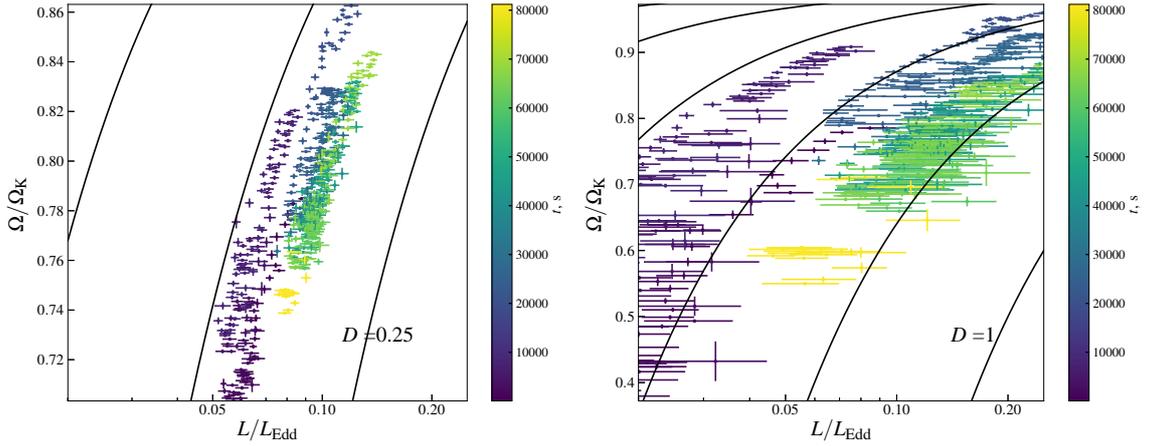}
 \caption{
Same as right panel of Fig.\,\ref{fig:OM} but for 
$D = 0.25$ ({\it left panel}) and $D=1$ ({\it right panel}).
} 
\label{fig:vardm}       
\end{figure*}

The variability spectrum of the mass accretion rate is encoded by two parameters, the root-mean-square variation of mass accretion rate logarithm $D$ and the slope of the power-law spectrum $p$. 
Their influence on the parallel tracks effect is shown in Figs.~\ref{fig:varp} and \ref{fig:vardm}. 
Redder variability spectrum allows the system to accrete longer at a steady rate different from the mean value, and thus increases the variations of mass and angular momentum. 
Thus, the parallel tracks effect is much more prominent for the case of red noise (right panel in Fig.~\ref{fig:varp}).
Harder variability spectrum ($p\lesssim 1$) makes the parallel tracks closer. 
However, the $L \propto \left(\Omega_{\rm K} - \Omega\right)^{-1}$ scaling still holds well. 

Different values of $D$ (see Fig.~\ref{fig:vardm}) also affect the prominence of the parallel tracks effect. 
As the amplitude of mass accretion rate variations increases by several times, the spacing between the short-term tracks increases from about 30\% to nearly two orders of magnitude. 

\section{Discussion}\label{sec:disc}

\subsection{Friction and depletion times}

As mentioned in the Introduction, the observed kHz QPO frequencies vary by a factor 1.5--2 in individual sources. While our model reproduces the parallel tracks effect in a broad range of parameters, strong variations in the rotation frequency of the BL appear only when the characteristic friction and mass depletion time scales are comparable. 
If friction is more efficient ($q\gtrsim 0.8$), the BL co-rotates with the star. 
If depletion is faster ($q\lesssim 0.5$), the BL co-rotates with the disk and loses angular momentum only with mass. 
Effectively, the second independent parameter necessary to reproduce the parallel tracks behavior exists only in a narrow range of $q$, meaning that there should be a physical reason for the depletion and friction time scales to be close to each other. 

\begin{figure}
\centering 
\includegraphics[width=1.0\columnwidth]{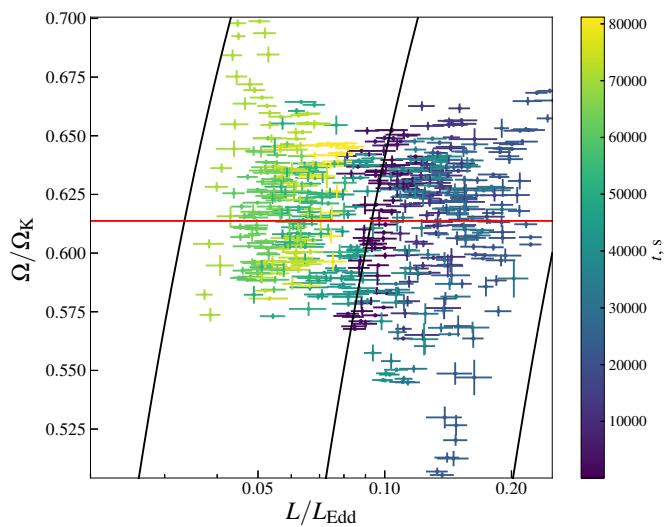}
 \caption{
Same as the right panel of Fig.~\ref{fig:OM} for a model with $\alpha = 10^{-7}$ and $q$ given by Eq.~(\ref{E:disc:q}). 
The horizontal red line corresponds to $\Omega_{\rm eq}$ given by Eq.~(\ref{E:disc:Oeq})
} 
\label{fig:qomega}       
\end{figure}

Such a similarity in the time scales may be explained if the BL is resolved in radial direction.
The radial flux of angular momentum consists of two parts, viscous $w_{r\varphi}R$ and advective $\omega R^2 \rho v$, where $v$ is vertical velocity,  $h\ll R$ is the height above the NS surface, $w_{r\varphi} = w_{r\varphi}(h)$ is the viscous stress component, and $\omega = \omega(h)$ is the rotation frequency, decreasing from $\Omega$ somewhere inside the BL to $\Omega_{\rm NS}$ at the NS surface. 
Because the viscous angular momentum transfer is directed outwards in the disk and inwards at the bottom of the BL, at some altitude it should be zero. 
Let us assume that $w_{\rm r\varphi}=0$ at the same altitude where $\omega = \Omega$, and write down angular momentum transfer along the radial coordinate
\begin{equation}\label{E:vphi}
    \ppardir{t}{\omega R^2} + v\ppardir{h}{\omega R^2} = - \frac{1}{R\rho}\ppardir{h}{w_{r\varphi}R^2}.
\end{equation}
In a steady-state case, $\rho v = \mbox{const}$, and the time derivative in Eq.~(\ref{E:vphi}) is zero. 
Integration yields
\begin{equation}
    \rho v \omega R^2 + w_{r\varphi}R = \rho v \Omega R^2.
\end{equation}
At the surface of the NS, $\omega(h) = \Omega_{\rm NS}$ and $w_{r\varphi}= W_{r\varphi}$, that implies
\begin{equation}
    \rho v \left( \Omega - \Omega_{\rm NS}\right) R^2 = R W_{r\varphi}.
\end{equation}
Multiplying this equation by $A$ and taking into account Eq.~(\ref{E:Tminus}) yields
\begin{equation}
    \frac{M}{t_{\rm depl}} \left( \Omega - \Omega_{\rm NS}\right) R^2 = \alpha g_{\rm eff} MR.
\end{equation}
Note that the mass flux $\rho v$ is related to the mass motion from the BL onto the surface of the star, hence we replaced $\rho v A$ with $M/t_{\rm depl}$. 
Substituting $g_{\rm eff}$ from Eq.~(\ref{E:geff}), we can express the $q$ parameter using Eq.~(\ref{E:q}) as
\begin{equation}\label{E:disc:q}
    q = \frac{\Omega_{\rm K} \left(\Omega- \Omega_{\rm NS}\right)}{\Omega_{\rm K}^2 - \Omega^2}.
\end{equation}
These estimates suggest that, instead of being an independent parameter, $q$ should depend on the rotation frequency of the BL. 
It is unclear if $q$ should change with the variations of $\Omega$. 
If $q$ depends on the mean or instantaneous value of $\Omega$, Eq.~(\ref{E:Oeq}) predicts an attractor for $\Omega/\Omega_{\rm K}$ and $q$.
Combining Eqs.~(\ref{E:Oeq}) and (\ref{E:disc:q}), we get  
\begin{equation}
\displaystyle q = \frac{2}{3+\Omega_{\rm NS}/\Omega_{\rm K}},
\end{equation}
and for the equilibrium rotation frequency
\begin{equation}\label{E:disc:Oeq}
    \Omega_{\rm eq} = \frac{\Omega_{\rm K}+\Omega_{\rm NS}}{2}.
\end{equation}
In Fig.~\ref{fig:qomega}, we show how our dynamical model behaves if the depletion time depends on rotation frequency as $t_{\rm depl} = \left( \Omega-\Omega_{\rm NS}\right) / \alpha \left( \Omega^2_{\rm K}-\Omega^2\right)$ for a fixed value of $\alpha$, that implies $q$ following Eq.~(\ref{E:disc:q}).
The parallel tracks effect is still reproduced in this version of the model. 

\subsection{Observable frequencies}

Here, we considered the rotation frequency of the BL as a characteristic QPO frequency. 
Though it is probably not the case, the real dynamical processes behind kHz QPOs are likely sensitive to $\Omega$. 
If the real oscillation frequencies are functions of $\Omega$ and $L$ or $M$, the parallel tracks effect is equally well reproduced, though the parameters of the correlation with radiation flux become different. 

In particular, for the Rossby-wave model considered in \citet{ANP20}, the characteristic oscillation frequencies are the epicyclic frequency 
\begin{equation}
    \Omega_{\rm e} \simeq 2\Omega\cos\theta,
\end{equation}
where $\theta$ is the co-latitude of the region where the oscillations are excited, and its aliases with rotation frequency, $\Omega_{\rm e}+n\Omega$, where $n$ is a whole number.
The oscillations are likely excited in the region of strongest latitudinal velocity shear, that is unstable to supersonic shear instability.
This naturally explains the multiplicity of kHz QPO frequencies and the difference between the frequencies that tends to be close to $\Omega_{\rm NS}$ (though not necessarily, see \citealt{mendez01}). 
Such a model also explains the characteristic values of the QPO frequencies and their correlation with the flux ($\cos\theta$ is likely a growing function of $L$, see \citealt{IS99,SP06}), and the different quality factors of the two QPO peaks (quality factors of the axisymmetric mode $n=0$ and all others should differ, as visibility effects enhance the periodic component in a non-axisymmetric case).
It is unclear, however, how to explain the existence of only two QPO peaks (probably, $n=0$ and $-1$). 
Higher harmonics may be below the sensitivity level, or their excitation conditions are different. 
If, instead of rotation frequency, we plot $\Omega_{\rm e}(L)$, the qualitative picture remains the same: tight correlation on the time scales about the time scales of the BL, that becomes worse on longer scales. 
The crucial point is the existence of the second variable, BL mass, slowly changing with time. 

In beat-frequency models of kHz QPO  \citep{miller01}, the higher peak corresponds to rotation frequency somewhere in the disk, and the lower -- to the beat between the higher frequency and stellar rotation. 
Both frequencies in such models change with a single variable parameter, the radius in the disk where the oscillations are excited. 
This radius apparently should change on the viscous timescale of the inner disk, and on longer times, the flux from the disk and the characteristic frequency should tightly correlate.
A way to reproduce a parallel-track picture in the framework of such a model is to add a contribution from the BL to the flux. 
The QPO frequency depends on the disk rather than total flux, and the dependence $\Omega(L)$ retains its slope but not the constant. 
Apparently, this is not the case, as the slope of the short-time relation between flux and frequency also changes considerably \citep{mendez99}, suggesting that the frequency itself is sensitive to the parameters of the BL rather than the disk. 

\section{Conclusions}\label{sec:conc}

We show that a very simple, zero-dimensional model of a BL accumulating mass and angular momentum from the disk allows to explain some of the properties of kHz QPOs. 
In particular, the model naturally reproduces the parallel tracks effect: the rotation frequency of the BL correlates with its luminosity at small time scales, but becomes uncorrelated at longer time scales. 

Such a `integrator' BL should have a distinct phase-lag signature: at high frequencies, its mass and rotation frequency should lag the variations of the mass accretion rate by $\Delta \varphi \simeq \uppi/2$. 
We expect the variations in kHz QPO frequencies in LMXBs to lag the variations of bolometric flux with a phase lag related to the contribution of the BL. 
Studying the cross-correlation properties of the kHz QPOs and flux variations in LMXBs will be an important test for the model and for our understanding of LMXBs in general.

\begin{acknowledgements}
This research was supported by the grant 14.W03.31.0021 of the Ministry of Science and Higher Education of the Russian Federation and the Academy of Finland grants 322779 and 333112. 
PA acknowledges support from the Program of Development of M.V. Lomonosov Moscow State University (Leading Scientific School `Physics of stars, relativistic objects and galaxies').
We thank the anonymous referee for the valuable comments. 
\end{acknowledgements}

\bibliographystyle{aa}
\bibliography{mybib}          
\label{lastpage}
\end{document}